\documentclass[twocolumn]{aastex701}
\usepackage{amsmath}
\usepackage{booktabs}

\begin{document}

\title{Beyond Disk Truncation: X-ray Reverberation Signatures of an Outflowing Corona}

\author[orcid=0000-0001-9206-1641]{Ke Qin}
\affiliation{School of Physics and Technology, Wuhan University, Wuhan 430072, People’s Republic of China;}
\email{}  

\author[orcid=0000-0002-8231-063X]{Bei You} 
\affiliation{School of Physics and Technology, Wuhan University, Wuhan 430072, People’s Republic of China;}
\email[show]{youbei@whu.edu.cn}

\author[orcid=0000-0002-5311-9078]{Adam Ingram} 
\affiliation{School of Mathematics, Statistics and Physics, Newcastle University, Herschel Building, Newcastle upon Tyne, NE1 7RU, UK;}
\email{}

\author[]{Barbara De Marco}
\affiliation{Departament de F\'isica, EEBE, Universitat Polit\`ecnica de Catalunya, Av. Eduard Maristany 16, Barcelona, E-08019, Spain;}
\affiliation{Institut d’Estudis Espacials de Catalunya, C. Esteve Terradas 1, Castelldefels, E-08860, Spain}
\email{}

\begin{abstract}

The corona in black hole X-ray binaries (BHXRBs) is likely dynamic during outbursts. A mildly relativistic outflowing corona has been proposed to explain hard spectra, weak disk reflection, and the higher than expected polarization degree observed in the hard state of X-ray binaries. In this work, we investigate the spectral and timing effects of coronal outflow using Monte Carlo radiative-transfer simulations. The model consists of a geometrically thin, optically thick truncated disk and an ellipsoidal corona with a prescribed bulk outflow velocity. We calculate the Comptonized continuum, the disk reflection component, and the lag-frequency spectra for different outflow velocities and disk truncation radii. We find that increasing the outflow velocity reduces the reflection fraction through relativistic beaming, because fewer Comptonized photons irradiate the disk. For $\beta \lesssim 0.5$, the high-frequency soft lag is only weakly affected by the outflow velocity, whereas increasing the disk truncation radius shifts the zero-crossing frequency($\nu_0$) to substantially lower frequencies. Thus, timing properties can help distinguish the coronal-outflow scenario from the disk-truncation scenario. We further apply the model to MAXI J1820+070 and find that its unusual $R-\Gamma$ anti-correlation during the plateau phase is qualitatively consistent with a contracting corona accompanied by increasing bulk velocity. Disk recession alone would predict the opposite evolution of $\nu_0$ under the assumptions of our model. Future polarization calculations will provide an additional test of extended outflowing-corona models.

\end{abstract}


\section{Introduction} 

Black hole X-ray binaries (BHXRBs) are among the most variable astrophysical sources, consisting of a stellar-mass black hole (BH) accreting material from a companion star. Most known BHXRBs are transient systems. They spend most of their time in quiescence but undergo luminous outbursts lasting from weeks to years \citep{2016ASSL..440...61B, 2016A&A...587A..61C}. During the outburst, BHXRBs exhibit distinct spectral and timing features driven by changes in accretion flow geometry and physical conditions \citep{2006ARA&A..44...49R, 2010LNP...794...53B, 2026Ap&SS.371...27L}. Therefore, BHXRBs provide a unique laboratory for studying accretion physics, disk-corona coupling, and the connection between accretion flows and relativistic jets.

A key feature is observed in the Hardness-Intensity Diagram (HID), which traces a characteristic $q$-shaped track during outbursts of BHXRBs \citep{2005A&A...440..207B, 2011BASI...39..409B}. At the onset of an outburst, sources typically enter the Low-Hard State (LHS), characterized by a hard X-ray spectrum (photon index $\Gamma \lesssim 2$) and lower luminosity. The X-ray emission in LHS is dominated by a non-thermal Compton component produced by hot electrons in the inner accretion flow, commonly referred to as the corona \citep{1993ApJ...413..680H, 1997ApJ...487..759D}. Low-energy seed photons from the accretion disk are inverse Compton scattered in the corona, producing a power-law continuum with a high-energy cutoff \citep{1979Natur.279..506S, 1998PhST...77...57P, 2009ApJ...690L..97P}.

The corona is thought to be spatially extended \citep{2008A&A...489..481K, 2021NatAs...5...94M, 2022Sci...378..650K, 2023Sci...381..961Y}. Measurements of disk emissivity profiles indicate that the illuminating region can extend over tens of gravitational radii, ($Rg=GM/c^2$) \citep{2012MNRAS.424.1284W, 2013MNRAS.429.2917F, 2015MNRAS.449..129W}. Since the innermost accretion region close to the compact object cannot be directly imaged, indirect probes are essential. X-ray reverberation mapping and polarization have become two powerful methods for constraining the spatial distribution and dynamics of the corona \citep{2019Natur.565..198K, 2021A&A...654A..14D, 2022Sci...378..650K, 2022MNRAS.516.5907M}.

X-ray time lags provide important information on the geometry and variability of the inner accretion flow. At low Fourier frequencies, hard lags are commonly observed, meaning that high-energy photons lag behind low-energy photons. These lags are usually interpreted as the result of inward propagation of mass accretion rate fluctuations \citep{2001MNRAS.327..799K, 2006MNRAS.367..801A, 2011MNRAS.415.2323I, 2023MNRAS.519.4434K}. At high Fourier frequencies, soft lags (softer photons lagging harder ones) are detected and generally attributed to reverberation: the light-travel time lag between direct coronal emission and photons that are Compton-scattered, irradiate the disk, reflect, and return to the observer \citep{2014MNRAS.439.3931E, 2014SSRv..183..453U, 2023MNRAS.520.5544G}. Although intrinsic continuum lags can dilute the measured lag amplitude, the zero-crossing frequency $\nu_0$, where the reverberation lag changes sign due to phase wrapping, is less sensitive to this dilution and can be used as a diagnostic of the inner accretion geometry \citep{2013MNRAS.430..247W, 2014SSRv..183..453U, 2014A&ARv..22...72U, 2018MNRAS.478..971M, 2021A&A...654A..14D}. In MAXI J1820+070, the observed evolution of $\nu_0$ has been interpreted as evidence for changes in the disk-corona geometry during the hard state \citep{2019Natur.565..198K, 2021ApJ...910L...3W, 2024MNRAS.528..558B, 2025ApJ...985..258Z}. However, the mapping between $\nu_0$ and a single characteristic geometrical scale may be energy dependent, since different reflection features can preferentially originate from different disk radii \citep{2026NatCo..17.2860Y}.

Most existing timing models assume that the corona is static. However, several observational results suggest that the corona may be dynamic and may carry a bulk outflow \citep{2026MNRAS.546ag261Z}. The reflection fraction ($R$), defined as the ratio of coronal emission illuminating the disk to that reaching the observer, varies significantly during BHXRB outbursts and is correlated with $\Gamma$ \citep{1999MNRAS.303L..11Z, 2023ApJ...945...65Y}. A mildly relativistic outflowing corona can naturally reduce the reflection fraction through relativistic beaming, because more photons are directed away from the disk and toward the outflow direction \citep{1999ApJ...510L.123B, 2001MNRAS.326..417M, 2021NatCo..12.1025Y, 2023ApJ...955...53F}. Multiwavelength correlations between X-ray and radio emission also support a connection between the corona and jet activity \citep{2013MNRAS.431L.107C, 2022MNRAS.512.2686Z, 2023A&A...679A..81K, 2025ApJ...994...54D}. In addition, bulk motions of corona may be a natural outcome of magnetic reconnection \citep{2025ApJ...979..199S}, and recent radiative general relativistic magnetohydrodynamic (GRMHD) simulations suggest that the Comptonizing region can be associated with a relativistic outflow along the jet sheath \citep{2024Ap&SS.369...68M, 2025ApJ...979..199S}. Such outflowing structures have also been investigated as a possible explanation for the high polarization degree measured in the hard state of X-ray binaries \citep{2022Sci...378..650K, 2023ApJ...949L..10P, 2025MNRAS.541.1774E}.

Although static corona models have been widely used to explain X-ray spectra and reverberation lags, the impact of coronal outflow on timing properties has not been systematically investigated. Previous studies have modeled either lamp-post coronae \citep{2014ApJ...782...76G, 2016ApJ...821L...1N, 2019MNRAS.488..324I} or extended coronae \citep{2016MNRAS.458..200W, 2025ApJ...985..258Z}, while dynamical outflowing corona models have mainly focused on spectral properties, especially the reflection fraction and spectral slope \citep{1999ApJ...510L.123B, 2001MNRAS.326..417M, 2013MNRAS.430.1694D, 2021NatCo..12.1025Y} or, more recently, X-ray polarization properties \citep{2023ApJ...949L..10P}. It is therefore important to examine whether timing observables, especially soft lags and $\nu_0$, can distinguish between changes in disk truncation and changes in coronal outflow velocity. This distinction is essential because both a receding inner disk and a faster outflowing corona can reduce the reflection fraction, but they predict different lag signatures.

In this work, we develop a Monte Carlo radiative transfer model for an accreting BH surrounded by a geometrically thin, optically thick truncated disk and an ellipsoidal outflowing corona. The model is based on the simulation framework of \cite{2018ApJ...858...82Y, 2020ApJ...897...27Y}, with the bulk motion of the corona included to study its effects on both the energy spectrum and timing properties. We then compare the predictions of the coronal outflow scenario with those of the disk truncation scenario, and apply the results to the unusual $R-\Gamma$ anti-correlation observed in MAXI J1820+070 during the plateau phase. This allows us to test whether the observed spectral evolution is more likely driven by changes in the inner disk radius or by the acceleration of an outflowing corona.

The structure of the paper is as follows. In Sect.\ref{sec:model}, we describe the physical model, including the disk-corona geometry, the implementation of coronal outflow, and the Monte Carlo simulation method. In Sect.\ref{sec:results}, we present the simulated energy spectra and timing results for different outflow velocities and disk truncation radii. In Sect.\ref{sec:discussion}, we discuss the implications for the $R-\Gamma$ relation and compare the disk truncation and coronal outflow interpretations, with a particular focus on MAXI J1820+070. Our main conclusions are summarized in Sect.\ref{sec:conclusions}.

\section{Model} \label{sec:model}

In this work, we use Monte Carlo radiative transfer simulations to follow thermal photons emitted from the disk, their inverse-Compton scattering in the corona, and the reflection component produced when Comptonized photons irradiate the disk. The simulation framework is based on the codes developed in \cite{2018ApJ...858...82Y, 2020ApJ...897...27Y} and further extended by \cite{2025ApJ...985..258Z}. In this work, we made modifications to the treatment of outflowing ellipsoidal corona and reflection compared to \cite{2025ApJ...985..258Z}.

\subsection{Geometry and dynamics \label{sec:geometry}}

The geometry of the corona in BHXRBs remains uncertain. Previous studies have considered a variety of configurations, including advection-dominated accretion flow (ADAF) within a truncated disk \citep{2006A&A...447..813F, 2008NewAR..51..733N}, slab corona \citep{1994ApJ...432L..95H, 2000MNRAS.318..180J}, lamp-post source \citep{2014ApJ...782...76G, 2016ApJ...821L...1N, 2021NatCo..12.1025Y}, vertically extended jet-base corona and more complex extended geometries \citep{2023A&A...679A..81K, 2025ApJ...986....3L, 2025ApJ...979..199S, 2025ApJ...985..258Z}. These models include different possible aspects of the disk-corona system, but the true coronal geometry is likely more complex \citep{2023MNRAS.525..854M, 2026MNRAS.551g1460M}. In this work, we adopt an ellipsoidal corona because it can represent both radially extended and vertically extended configurations in a simple parametric form. An ellipsoidal corona is more in line with recent results from X-ray polarimetry,  which generally favour low aspect ratio geometries \citep{2022Sci...378..650K, 2024ApJ...968...76I}.

\begin{figure}[h]
    \centering
    \includegraphics[width=\linewidth]{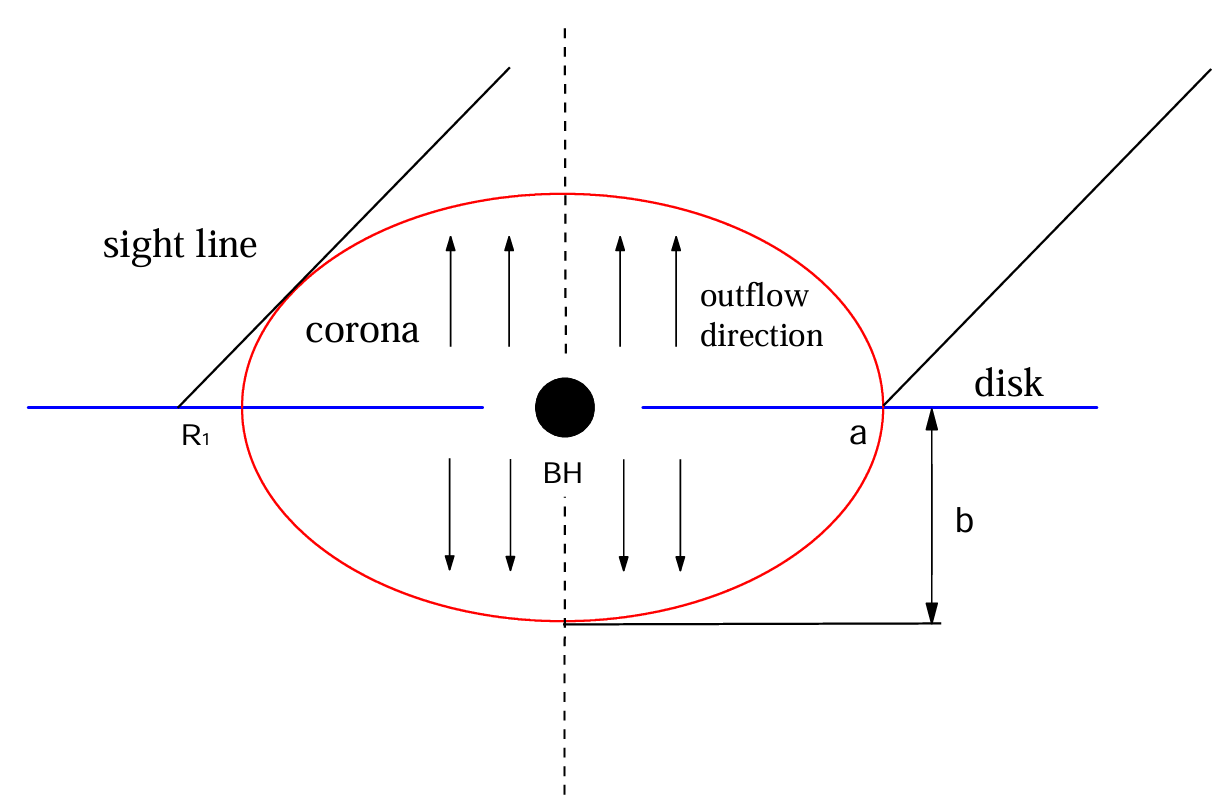}
    \caption{The cross-section of ellipsoidal corona. The red solid line represents the boundary of corona, the blue solid line represents the disc, and the black solid line represents the sight line of observers. The maximum radius of intersection point of the sight line with disk is $R_1$. The upward arrow indicates the outflowing direction of corona.}
    \label{fig:modelgeo}
\end{figure}

We consider a geometrically thin and optically thick accretion disk located in the $xy$ plane. The disk is truncated at an inner radius $R_{\rm tr}$ and extends outward to $R_{\rm max}$. In the simulations presented below, $R_{\rm max}$ is uniformly set to $1000\,Rg$. The BH is located at the origin and is surrounded by an ellipsoidal corona (see Fig. \ref{fig:modelgeo}). The shape of the corona can be parameterized by the following equation:
\begin{equation}
\frac{x^2+y^2}{a^2}+\frac{z^2}{b^2}=1,  \label{eq:hyperboloid}
\end{equation}
where $a$ is the semi-axis in the disk plane and $b$ is the semi-axis along the vertical direction. When $a>b$, the corona is more radially extended, similar to a hot atmosphere above the disk. When $b>a$, the corona is more vertically extended, resembling the base of a jet. Unless otherwise stated, the coronal density and electron temperature are assumed to be spatially uniform.

\subsection{Monte Carlo radiative transfer and variability\label{sec:MC}}

The hard-state X-ray spectrum of BHXRBs is generally dominated by a Comptonized continuum with a high-energy cutoff, together with thermal disk emission and reflection. We therefore include three radiative components in the simulation: thermal radiation from the thin disk, inverse-Compton scattering in the corona, and reflection from the disk surface.

The seed photons are generated from the accretion disk. The radial emissivity of the disk is calculated using the Novikov-Thorne model \citep{1973blho.conf..343N}. The initial emission radius of each photon is sampled from the cumulative distribution of the disk flux, following the method used in \cite{2018ApJ...858...82Y}. The initial direction of photons is isotropic. If a disk photon enters the corona, it can undergo inverse-Compton scattering by electrons. The scattering process is calculated following the standard Monte Carlo treatment of Comptonization \citep{1983ASPRv...2..189P, 1984AcA....34..141G}. The optical depth, electron density, and electron temperature determine the scattering probability and the energy exchange between photons and electrons.

The photon trajectory is followed until the photon either escapes to infinity or strikes the disk. As in \cite{2025ApJ...985..258Z}, We calculate photon propagation in the Newtonian framework, neglecting general relativistic effects such as light bending. However, special relativistic effects caused by the bulk motion of the corona are included in the Comptonization process. This approximation is adequate for the present purpose because our main goal is to isolate the effect of coronal outflow on the relative changes in the spectrum and time lags. General relativistic light bending may affect the absolute value of reflection fraction when the emitting region is very close to the BH, but it is not expected to change the qualitative trends studied here.

To simulate fast variability, we introduce propagating fluctuations in the mass accretion rate of the disk. The accretion flow is divided into 20 logarithmically spaced radial bins, each centered at $r_n$. Following \cite{2011MNRAS.415.2323I}, the local variability generated at each radius is described by a zero-centered Lorentzian power spectrum:
\begin{equation}
    |{\dot M}(r_n,f)|^2 \propto \frac{1}{1+(f/f_{\mathrm{visc}}(r_n))^2},
    \label{eq:periodogram}
\end{equation} 
\noindent where $f$ is the Fourier frequency. $f_{\mathrm{visc}}(r_n)=f_{\rm max}(r_n/r_{\rm tr})^{-3/2}$ represents the viscous frequency at radius $r_n$, where $f_{\rm max}$ is a free parameter, indicating the maximum viscous frequency at $R_{\rm tr}$. Then, we perform the inverse Fourier transform on Equation(\ref{eq:periodogram}) using the randomization method proposed by \cite{1995A&A...300..707T} to generate the local accretion rate $\dot m(r_n,t)$ at radius $r_n$. We treat each radial bin of the disk as a fluctuanting source. These fluctuations propagate inward and couple multiplicatively with fluctuations generated at smaller radii \citep{2013MNRAS.434.1476I, 2026ApJ..1006..177H}. Therefore, the total fluctuations at the radius $r_n$ and time $t$ is the superposition of the local fluctuations before time $\Delta t_{ln}$:

\begin{equation}
    \dot m_{\mathrm{total}}(r_n,t)=\prod_{l=1}^n \dot m(r_l,t-\Delta t_{ln})
\end{equation}

\noindent where $\Delta t_{ln}$ is the propagation lag between radius $r_l$ and $r_n$, which can be expressed by
\begin{equation}
    \Delta t_{ln}=\sum_{q=l+1}^n\frac{\mathrm d r_q}{r_q f_{\mathrm{visc}(r_q)}}
\end{equation}
\noindent and $\mathrm d r_n=r_n-r_{n-1}$.

In the Monte Carlo calculation, the variability is introduced by assigning a time-dependent weight to each photon according to the accretion-rate fluctuation at its emission radius and emission time. The observed light curve in a given energy band is then obtained by summing the weights of photons that reach the observer, and these photons include the information of the viscous propagation lag and light-travel lag (including the light travel time between successive scattering orders and the reverberation delay between directly observed and reflected photons). This treatment allows the model to include low-frequency hard lags associated with inward propagation of accretion rate fluctuations and high-frequency soft lags associated with reverberation lags. For simplicity, we do not include additional fluctuations generated inside the corona itself, and we assume a uniform optical depth and temperature throughout the corona. The coronal temperature and optical depth are treated as phenomenological parameters, rather than being derived from a self-consistent heating-cooling balance. The time lags in our simulations are not produced by variations of the photon index. Instead, the variability is introduced through fluctuations of the seed photon luminosity, which only modify the normalization of the Comptonized spectrum without changing the electron temperature or photon index.

\subsection{Implementation of coronal outflow\label{sec:outflow}}

The static corona approximation may not be sufficient for BHXRBs. An outflowing corona was proposed to explain weak disk reflection and hard X-ray spectra, because relativistic beaming can reduce the fraction of coronal photons that illuminate the disk \citep{1999ApJ...510L.123B, 2001MNRAS.326..417M}. Similar ideas have been explored for slab coronae \citep{2000MNRAS.318..180J}, lamp post sources \citep{2021NatCo..12.1025Y}, and jet-base geometries \citep{2023A&A...679A..81K, 2025ApJ...979..199S, 2025ApJ...993..167M}. More recently, bulk outflowing corona models have also been proposed to explain the higher than expected polarization degree observed from X-ray binaries in the hard state \citep{2022Sci...378..650K, 2023ApJ...949L..10P, 2025MNRAS.541.1774E}. However, these studies mainly focused on polarization properties and did not investigate timing signatures.

In this work, we prescribe a bulk outflow velocity $v=\beta c$ for the coronal plasma. The outflow is directed along the $z$-axis, perpendicular to the disk plane and away from the BH, and $\beta$ is assumed to be spatially uniform throughout the corona. For a photon propagating inside the outflowing corona, its energy and propagation direction are first Lorentz-transformed from the BH rest frame to the comoving frame of the corona. The thermal velocity of the scattering electron is then sampled in the coronal comoving frame, and the inverse-Compton scattering is calculated using the same Monte Carlo prescription as in the static case. The energy and propagation direction of the scattered photon obtained in the comoving frame are subsequently Lorentz-transformed back to the BH rest frame, in which the photon trajectory and propagation time are followed.

In principle, the differential optical depth in a moving medium is direction dependent. For a coronal proper electron density $n'$, it can be written as $d\tau = n' \Gamma_b (1 - \beta \mu) \sigma_{KN} (E') \, dl$\citep{2014MNRAS.440.3292L}, where $\Gamma_b = (1 - \beta^2)^{-1/2}$ is the bulk Lorentz factor, $\mu$ is the cosine of the angle between the photon propagation direction and the bulk velocity, and $E'$ is the photon energy measured in the coronal comoving frame. In the present calculations, we do not explicitly include this directional correction to the optical depth. Instead, the coronal optical depth is treated as an effective free parameter, together with the electron temperature, controlling the Comptonization strength and hence the photon index of the emergent spectrum. For a given outflow velocity, these spectral parameters can be adjusted to reproduce a physically reasonable photon index. Therefore, including the velocity-dependent correction to the optical depth would mainly require a corresponding re-adjustment of the effective optical depth when matching the continuum slope. Since the primary aim of this work is to investigate the relative timing and reverberation signatures associated with coronal outflow rather than to determine the coronal density or optical depth self-consistently, we adopt the above simplified treatment. We note that extremely large outflow velocities or optical depths may not be physically realistic, because they would imply an excessively large mass outflow rate that could exceed the available accretion power and become strongly super-Eddington. The parameter ranges explored here are therefore intended to investigate the relative effects of coronal outflow on spectral and timing properties, rather than represent a complete set of physically self-consistent outflow solutions.

\subsection{Calculation of reflection\label{sec:reflection}}

A fraction of the Comptonized photons escaping from the corona irradiates the disk and is reprocessed through photoelectric absorption, fluorescent line emission, Compton scattering, and bremsstrahlung radiation \citep{2005MNRAS.358..211R}. The resulting reflection spectrum contains characteristic features such as the Fe K$\alpha$ line and the Compton hump \citep{1991MNRAS.249..352G}. Directly simulating all atomic processes in the disk atmosphere is computationally expensive. We therefore use the \textsc{XILLVER} table model to calculate the non-relativistic reflection spectrum \citep{2013ApJ...768..146G, 2013MNRAS.430.1694D}.

For each simulation, we collect the Comptonized photons that strike the disk and use them to estimate the incident spectrum in different radial bins. The $\Gamma$ of the incident spectrum and the incident flux normalization are then used as inputs for the \textsc{XILLVER} interpolation. The \textsc{XILLVER} table is normalized through the incident flux condition \citep{2016A&A...590A..76D}:

\begin{equation}
    \int^{1000\mathrm{keV}}_{0.1\mathrm{keV}}\mathcal F_X(E)\mathrm{d}E=10^{-20}\frac{n_0\xi_0}{4\pi}
    \label{eq:norm},
\end{equation}

where the density and ionization parameter in the XILLVER normalization are fixed to the reference values $n_0=10^{15}\mathrm{~cm^{-3}}$ and $\xi_0=1\mathrm{~erg~cm~s^{-1}}$, respectively. These values define the normalization convention of the XILLVER table and are not the input disk parameters adopted in our simulations. The photon index $\Gamma$ and actual incident flux used for the reflection calculation are determined from the Monte Carlo simulation and are used for interpolation and rescaling of the XILLVER spectrum. Other input parameters include the inclination angle $i$, the ionization parameter $\xi$, and the iron abundance $A_{\rm Fe}$.

Additionally, we need to consider the attenuation effect of coronal obscuration on the reflective components from the inner disk. \cite{2025ApJ...985..258Z} adopted an approximate two-zone method to describe the shielding of disk reflection, which is a simplified treatment. In this work, we define the probability that disk photons generated at different radii and different polar angles reach infinity without undergoing scattering as the coronal transmittance. Its calculation is given by the following:

\begin{equation}
    p_{\rm esc}(r,i) = \frac{N_{\rm esc}(r,i)}{N_{\rm gen}(r,i)},
    \label{eq:pesc}
\end{equation}

\noindent where $N_{\rm gen}(r,i)$ represents the number of photons that are emitted from the disk at radius $r$ and polar angle $i$, while $N_{\rm esc}(r,i)$ is the number of photons that reach infinity without scattering. For an ellipsoidal corona geometry, the sight lines of observer create a shadow projection on the disk, as shown in Fig.\ref{fig:modelgeo}. The farthest position of projection is at $R_1$, and the nearest position is at parameter $a$. The reflected components within the projected area will be attenuated by the transmission rate $p_{esc}(r,i)$ before reaching the observer. We adopted an approximate method to calculate the shielding effect at different radii:

\begin{equation}
N_{\rm refl,obs}(r,i)=N_{\rm refl,in}(r,i) \times p_{\rm esc}(r,i),
\label{eq:shield}
\end{equation}

\noindent where $N_{\rm refl,in}(r,i)$ is the intrinsic reflected photon number before attenuation. The total observed reflection component is obtained by summing the attenuated reflected photons over all disk radii:\( N_{refl,all}(i) = \sum_r N(r,i)_{refl,obs} \)

In Equation (\ref{eq:shield}), only reflected photons that escape through the corona without further scattering are included in the observed reflection component. Reflected photons that undergo Compton scattering in the corona are not followed further. Previous studies have shown that such scattering strongly smooths the reflection features and causes the scattered component to blend into the primary Comptonized continuum \citep{2001MNRAS.326..417M, 2001MNRAS.328..501P, 2015MNRAS.448..703W}. The scattered reflection component is expected to modify the intrinsic hard lags and contribute to the dilution of the soft lags. Recent modeling of the BHXRB EXO 1846-031 also found a reflected-photon scattering fraction of order $10\%$, while including this component produced only minor changes in the main fitted parameters \citep{2024PhRvD.110d3021L}. We therefore neglect this secondary Comptonized-reflection component in the present simulations to reduce computational complexity.

\section{Results} \label{sec:results}

Our main goal is to quantify how the bulk velocity of the corona affects the Comptonized continuum, the disk reflection component, and the reverberation lag. Unless otherwise stated, we adopt the fiducial parameters listed in Table \ref{tab:paramenters}. These parameters define the size of the ellipsoidal corona and disk, the mass accretion rate and ionization of disk, the maximum viscous frequency, and the observer inclination. The mass of BH is set at $7.6 M_\odot$, which is consistent with the estimated value of MAXI J1820+070 \citep{2021A&A...654A..14D}. The simulation time scale is 200 seconds, with a time resolution of 1 millisecond. We first examine the time-averaged energy spectra and the corresponding reflection fraction, and then investigate the lag-frequency spectra.

\begin{table*}[htbp]
\centering
\caption{Some key parameters in this model and their definitions}\label{tab:paramenters}
\begin{tabular}{@{}lclllcl@{}}
\toprule
 & Notation &  & Definition & &Default value& Unit \\ \midrule
 & $a$ & & Semi-axis length that intersects the $xy$ plane & & 50 & $Rg$ \\
 & $b$ & & Semi-axis length that intersects the $z$-axis & & 40 & $Rg$ \\
 & $M_{BH}$  & &  The mass of BH & & 7.6 & $M_\odot$ \\
 & $\dot{M}$  & &  Mass accretion rate  & & 0.01 & $\dot{M}_{Edd}$ \\
 & $R_{tr}$ & & Truncation radius of disk & & 10 & $Rg$ \\
 & $R_{max}$ & & Maximum radius of disk & & 1000 & $Rg$ \\
 & $f_{max}$ & & Maximum viscous frequency of disk & & 4 & Hz \\
 & $log\;\xi$ & & Ionization parameter of disk & & 3.0 &  $erg\;cm\;s^{-1}$\\
 & $i$ & & Inclination angle with respect to the normal to the disk & & 60 & $^\circ$ \\ \bottomrule
\end{tabular}
\end{table*}

\begin{figure}[h]
    \centering
    \includegraphics[width=\linewidth]{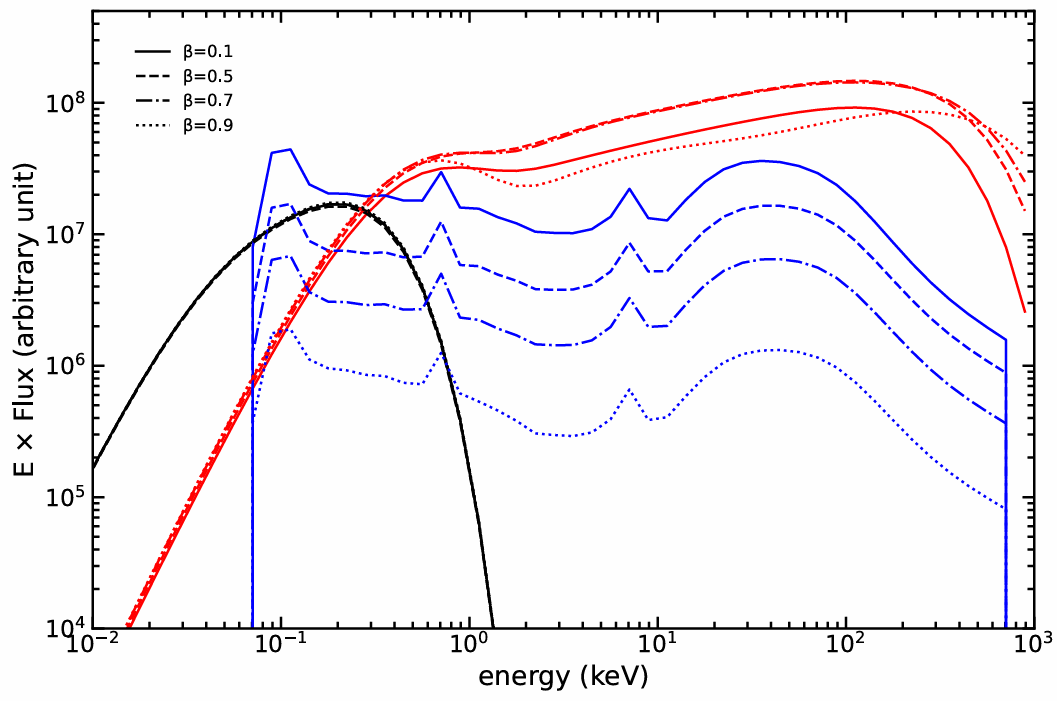}
    \caption{The time-averaged energy spectra for different $\beta$. The $Y$-axis is labeled by the energy multiplied by the flux. We do not show the $\beta=0$ case because it overlaps with the $\beta=0.1$ case. The black lines represent the disk component, the red lines represent the Compton component, and the blue lines represent the reflection.}
    \label{fig:spectrum}
\end{figure}

Fig.\ref{fig:spectrum} shows the time-averaged energy spectra for different outflow velocities ($\beta$). As $\beta$ increases, the reflection component becomes progressively weaker. For low outflow velocities, especially $\beta \lesssim 0.5$, this suppression is modest. The spectrum for $\beta=0.1$ is almost identical to that of a static corona, and therefore the $\beta=0$ case is not shown for visual clarity. When the outflow becomes faster, the reflection component is strongly reduced. This trend demonstrates that a mildly relativistic coronal outflow can significantly change the angular distribution of the scattered photons. For $\beta \gtrsim 0.7$, the Comptonized continuum dominates over the reflection component by more than an order of magnitude in most energy bands, which means that when the outflow is highly relativistic, reflection features such as the Fe K line (6-7 keV) and the Compton hump (30-70 keV) may not be observable in the spectrum. 

\begin{figure}[h]
    \centering
    \includegraphics[width=\linewidth]{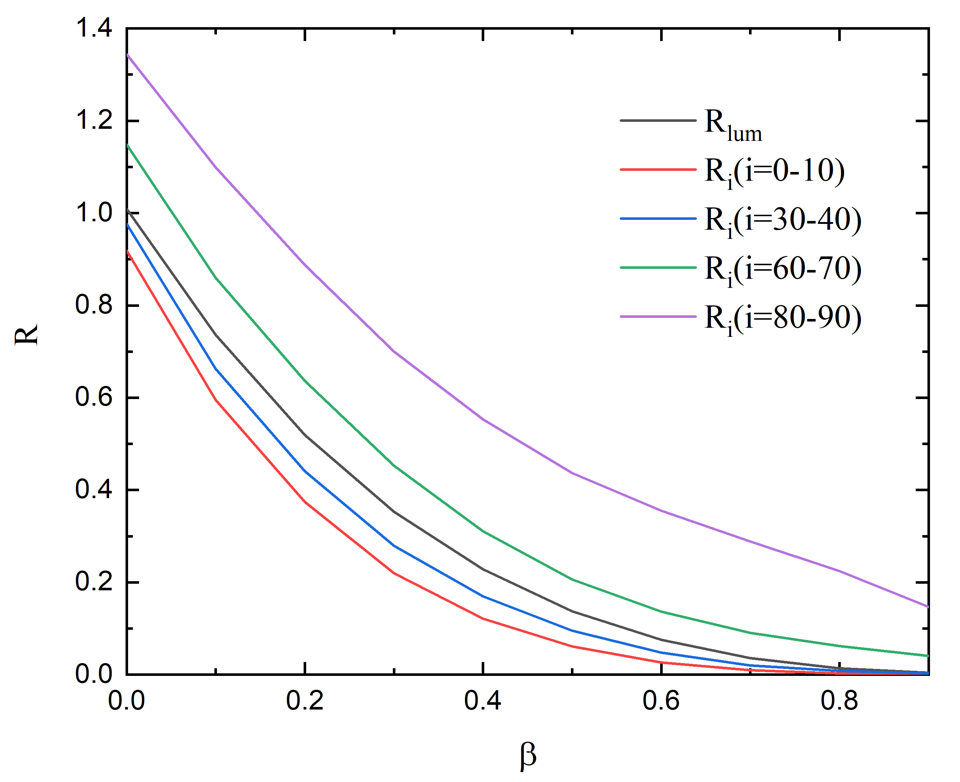}
    \caption{The reflection fraction as a function of $\beta$. The definition represented by the black line is based on Equation \ref{eq:R_lum}. The definition represented by colored lines is based on Equation \ref{eq:Ri}, which is related to the inclination angles ($i$).}
    \label{fig:R}
\end{figure}

To further quantify the beaming effect, we show the reflection fraction as a function of $\beta$ in Fig.\ref{fig:R}. We first define a luminosity-based reflection fraction \citep{1999MNRAS.303L..11Z}:

\begin{equation}
    R_{\rm lum}=\frac{L_{\rm inc}}{L_{\inf}},
    \label{eq:R_lum}
\end{equation}

\noindent where $L_{\rm inc}$ is the Comptonized luminosity incident on the disk and $L_{\rm inf}$ is the Comptonized luminosity escaping to infinity. Here, $L_{\rm inc}$ and $L_{\rm inf}$ are calculated by integrating the photon energy flux over the full energy range covered by our simulations (from $10^{-3}$ keV to $10^{3}$ keV). Since the adopted energy range covers the dominant part of the Comptonized emission, these quantities can be regarded as approximately bolometric luminosities. $R_{\rm lum}$ measures the global redistribution of photons between the disk and infinity. The black curve in Fig.\ref{fig:R} shows that $R_{\rm lum}$ decreases rapidly with increasing $\beta$. This decrease is a direct consequence of relativistic beaming: as the corona moves away from the disk, a larger fraction of the scattered photons is directed along the outflow direction, while fewer photons return to illuminate the disk \citep{2016MNRAS.458..200W, 2021NatCo..12.1025Y}. Therefore, the weakening of reflection spectrum in Fig.\ref{fig:spectrum} is consistent with the decline of reflection fraction in Fig.\ref{fig:R}.

We also calculate the inclination-dependent reflection fraction introduced by \cite{2016A&A...590A..76D},

\begin{equation}
    R_i = \frac{F_{\rm inc}}{F_{\rm obs}(i)},
    \label{eq:Ri}
\end{equation}

\noindent where $F_{\rm inc}$ is the flux incident on the disk and $F_{\rm obs}(i)$ is the observed Comptonized flux at inclination angle $i$. Both $F_{\rm inc}$ and $F_{\rm obs}(i)$ are obtained by integrating the energy flux over the full simulated range of 0.001-1000 keV. This definition is also known as the reflection amplitude \citep{1999ApJ...510L.123B, 2000MNRAS.318..180J}, and it is more closely related to the reflection fraction inferred from spectral fitting, because the observed continuum depends on viewing angle when the primary source is anisotropic. The colored curves in Fig.\ref{fig:R} show $R_i$ for different inclination angles. The decline of $R_i$ with $\beta$ is steeper at smaller inclination angles, because observers closer to the outflow axis receive a more strongly beamed Comptonized continuum. This behavior is consistent with previous simulations of outflowing coronae, in which relativistic beaming reduces the disk-illuminating flux and enhances the flux emitted toward the outflow direction \citep{1999ApJ...510L.123B, 2000MNRAS.318..180J}.

\begin{figure}[h]
    \centering
    \includegraphics[width=\linewidth]{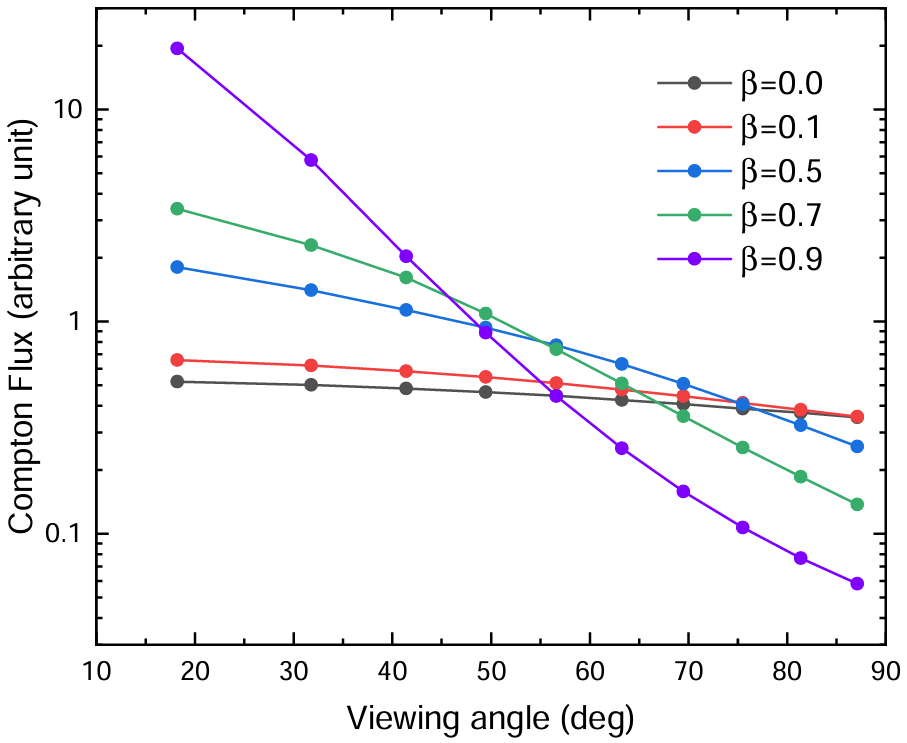}
    \caption{The Compton flux received by the observer at different angles. Different colored lines represent different values of $\beta$.}
    \label{fig:flux}
\end{figure}

According to the beaming effect, as the coronal outflow velocity increases, the flux incident on the disk decreases while the flux emitted toward infinity increases. However, in the energy spectra shown in Fig.\ref{fig:spectrum}, the Comptonized flux does not exhibit a significant change. This is mainly due to the specific viewing angle adopted in those simulations. The angular redistribution of the Comptonized photons is shown more directly in Fig.\ref{fig:flux}. As $\beta$ increases, the emission becomes increasingly concentrated toward small inclination angles, i.e., along the outflow direction. At an edge-on view, the Comptonized flux even decreases as $\beta$ increases.

X-ray timing analysis typically requires performing the Fourier transform on the light curve to reveal features at different frequencies. We use the Python package STINGRAY \citep{2019ApJ...881...39H} to calculate the Fourier cross-spectrum of the light curves in selected energy bands. The $200\,s$ simulated light curves have a time resolution of $1\,ms$ and are divided into $1\,s$ segments for the Fourier analysis. The cross spectra from individual segments are averaged to obtain the final cross spectrum, and the lag-frequency spectra are logarithmically rebinned with a rebinning factor of $f=0.4$. The time lag $\tau(\nu)$ related to frequency between the selected energy bands is given by $\tau(\nu) = \phi(\nu) / (2\pi\nu)$, where $\nu$ is the Fourier frequency, and $\phi(\nu)$ is the phase lag of cross-spectrum \citep{1999ApJ...510..874N, 2021A&A...654A..14D}. 

\begin{figure}[h]
    \centering
    \includegraphics[width=\linewidth]{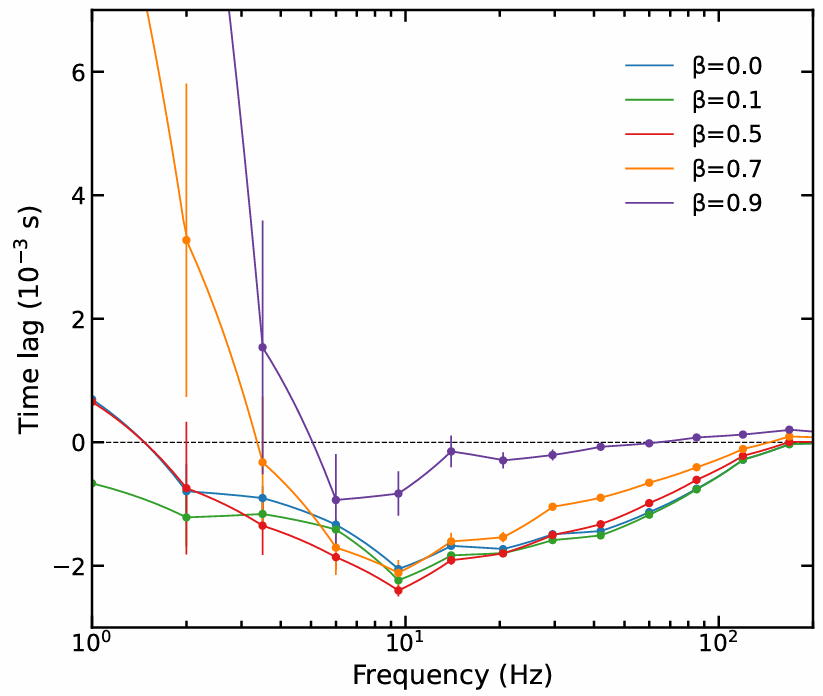}
    \caption{The lag-frequency spectra for different $\beta$, calculated between the energy bands of 0.1-0.3 keV and 2.0-5.0 keV.}
    \label{fig:lag}
\end{figure}

In Fig.\ref{fig:lag}, we present the lag-frequency spectra for different values of $\beta$. The selected energy band is 0.1 - 0.3 keV, dominated by the reflection, and 2.0 - 5.0 keV, dominated by the Compton component (see Fig.\ref{fig:spectrum}). The lag-frequency spectra show the hard lag at low frequency and the soft lag at high frequency, which is consistent with the observations \citep{2019Natur.565..198K, 2021A&A...654A..14D}. For relatively low velocities $\beta \lesssim 0.5$, the zero-crossing frequency is $\nu_0\sim160$ Hz, the lag spectra show only minor variations, indicating that the reverberation signal is largely preserved. As $\beta$ increases, the soft-lag amplitude is progressively suppressed, and for sufficiently large velocities, the reverberation signature becomes significantly weakened. When $\beta=0.9$, $\nu_0\sim60$ Hz. Since the disk-corona geometry is fixed in these simulations, the intrinsic light-travel distance remains unchanged, and the observed reduction in soft-lag amplitude should therefore be attributed to the decreasing contribution of the reflected emission rather than to any geometrical contraction. This result indicates that coronal outflow can substantially modify timing observables and must be taken into account when interpreting reverberation lags in terms of disk-corona geometry.

\begin{figure}[h]
    \centering
    \includegraphics[width=\linewidth]{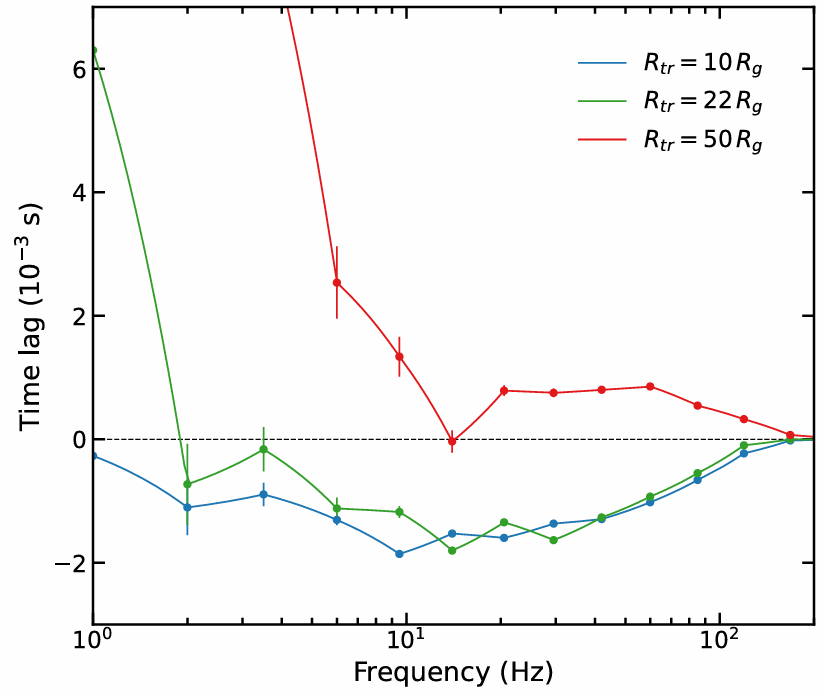}
    \caption{The lag-frequency spectra for $\beta=0$ with different disk truncation radii ($R_{\rm tr}$). The cases with $R_{\rm tr}=10$, $22$, and $50~R_{\rm g}$ have reflection fractions of approximately 1.0, 0.75, and 0.2, respectively, corresponding to the reflection fractions of the $\beta=0$, $0.1$, and $0.5$ cases in Fig.~\ref{fig:lag}. The photon index $\Gamma$ is also fixed to be the same as that adopted in Fig.~\ref{fig:lag}. The different timing behaviors provide a diagnostic to distinguish disk truncation from coronal outflow scenarios.}
    \label{fig:lagRtr}
\end{figure}

Although the reflection fraction-photon index ($R-\Gamma$) relation provides important information about the evolution of the disk-corona system, it alone cannot distinguish between a changing coronal outflow velocity and a changing disk truncation radius, since both scenarios can reduce the observed reflection fraction. To illustrate the difference between these two possibilities, in Fig.\ref{fig:lagRtr}, we further show the timing properties of a static corona with different disk truncation radii. The $\Gamma$ is kept identical to the corresponding outflowing-corona cases in Fig.\ref{fig:lag}. For the static-corona cases, we adopt $R_{\rm tr}=10$, $22$, and $50\,Rg$, which produce reflection fractions of approximately 1.0, 0.75, and 0.20, respectively. These values correspond approximately to the cases of $\beta=0.0$, $0.1$, and $0.5$ in Fig.\ref{fig:lag}. As the $R_{\rm tr}$ increases, the intrinsic disk-corona light travel distance becomes longer, leading to a stronger reverberation signal. However, the reduced reflection contribution also enhances the dilution effect, which suppresses the observed soft lag amplitude. When $R_{\rm tr}=50\,Rg$, the soft lag signal disappears due to dilution. Therefore, the timing evolution in the truncation-radius scenario differs from that produced by increasing outflow velocity (e.g. compare the red curves in Fig.\ref{fig:lag} and \ref{fig:lagRtr}), providing an additional diagnostic beyond spectral fitting. A detailed comparison between these two scenarios is presented in Sect.\ref{sec:correlation}.

\section{Discussion} \label{sec:discussion}

\subsection{The coronal temperature\label{sec:temperature}}

The electron temperature of the corona is an important parameter because it determines the energy gain of photons during inverse-Compton scattering and therefore affects the photon index $\Gamma$. The spectral slope of thermal Comptonization is jointly determined by the electron temperature and optical depth of the corona. In our model, both quantities are treated as phenomenological parameters rather than being calculated self-consistently from the heating-cooling balance of the plasma \citep{2025MNRAS.536.3284U}. Physically, the balance between plasma heating and radiative cooling can change when the corona becomes dynamic. As the outflow velocity increases, relativistic beaming modifies the angular distribution of scattered photons, allowing photons to escape more efficiently along the outflow direction and reducing the effective scattering path length. If the coronal temperature is kept fixed, photons undergo fewer scatterings and gain less energy, leading to a softer Comptonized spectrum.

\begin{figure}[h]
    \centering
    \includegraphics[width=\linewidth]{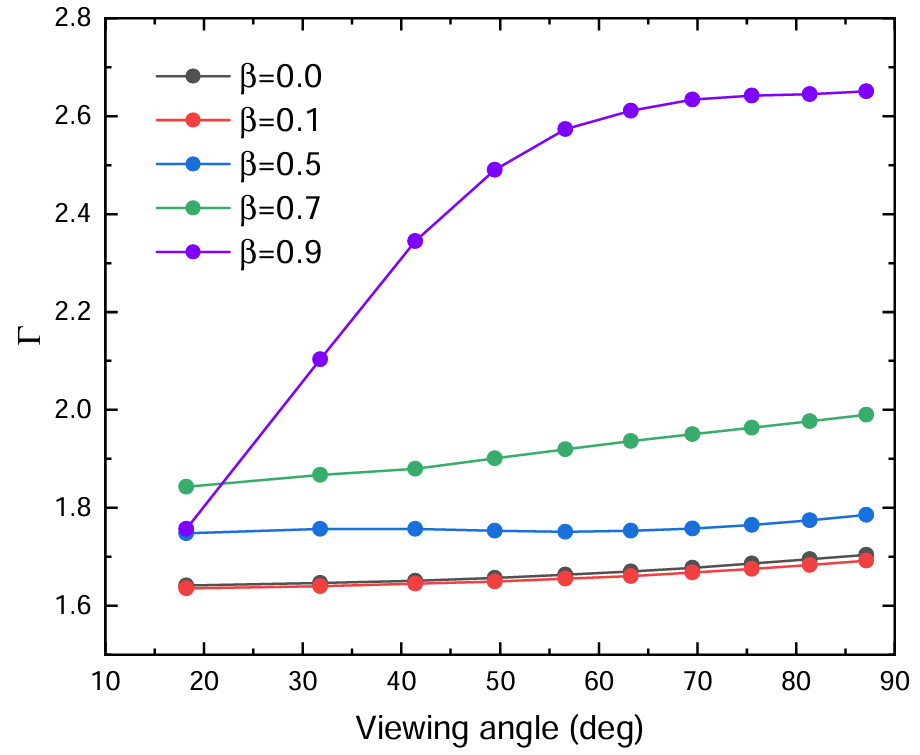}
    \caption{$\Gamma$ as a function of viewing angles for different $\beta$ when the coronal temperature is fixed at 100 keV.}
    \label{fig:gamma}
\end{figure}

This effect is illustrated in Fig.\ref{fig:gamma}, where we show the results obtained with a fixed coronal temperature. In this case, $\Gamma$ increases rapidly with $\beta$, especially for $\beta \gtrsim 0.7$, and becomes larger than typically observed in the hard state. This behavior is more pronounced at high inclination angles, where photons tend to experience fewer scatterings in a fast outflow. To maintain a physically reasonable spectral slope, we therefore increase the coronal temperature with $\beta$ in our main simulations. This adjustment compensates for the reduced scattering efficiency and ensures that the Comptonized spectrum remains consistent with observations.

This treatment is also supported by observational evidence. In many BHXRBs, the reflection fraction $R$ is positively correlated with $\Gamma$, indicating that a lower $R$ is generally associated with a harder spectrum \citep{2010LNP...794...17G,2023RAA....23g5005D, 2023ApJ...945...65Y}. If the decrease of $R$ were driven purely by the outflow velocity increasing, with the coronal temperature remaining constant, $R$ would instead anti-correlate with $\Gamma$ (see Fig.\ref{fig:gamma}). However, the observed positive correlation between $R$ and $\Gamma$ can be reproduced if the coronal temperature increases with increasing $\beta$. In our simulations, this temperature variation is not derived from a self-consistent coronal energy balance, but is treated as a phenomenological prescription to explore whether an outflowing corona can reproduce the observed spectral evolution. No specific mathematical relation between the coronal temperature and $\beta$ is assumed, instead, the coronal temperature is adjusted for different $\beta$ to obtain $\Gamma$ consistent with observations. A self-consistent dependence of the electron temperature on the $\beta$ requires a detailed treatment of coronal heating, radiative cooling, and disk-corona feedback, which is beyond the scope of this work. Observational studies of BHXRBs have shown that the coronal temperature can vary significantly during spectral evolution \citep{2015ApJ...813...84G, 2023ApJ...945...65Y, 2025MNRAS.538.1143L, 2026ApJ..1000...20Y}, indicating that changes in the thermal state of the corona should be considered when interpreting the $R-\Gamma$ relation.

\subsection{Reflection Fraction\label{sec:R}}

The reflection fraction is sensitive to the angular distribution of the primary X-ray emission. For an isotropic point-like X-ray source above an infinitely extended disk, the inclination-dependent reflection fraction $R_i$ is expected to be close to unity in the non-relativistic limit \citep{2016A&A...590A..76D}. In our simulations, however, $R_i$ is not exactly equal to 1 even when $\beta=0$, as shown in Fig.\ref{fig:R}. There are two main reasons for this behavior. First, the disk in our model has an inner truncation radius $R_{\rm tr}$ and an outer boundary $R_{\rm max}$. Therefore, not all photons emitted downward from the corona can be intercepted by the disk. Second, even for a static corona, the Comptonized emission is not perfectly isotropic. The seed photons are emitted from the disk, so the first-order scattering is affected by the anisotropic angular distribution of the incident seed radiation \citep{1995ApJ...449L..13S, 1996A&AS..120C.475S, 1997ApJ...476..620H}, which would result in an anisotropic total Compton component.

Relativistic effects can also modify the reflection fraction by bending photon trajectories toward the disk. This effect is important when the corona is located very close to the BH or when the disk extends to a small radius. However, for the typical parameters adopted in our simulations, the corona extends to tens of $Rg$ and the disk is truncated at $R_{\rm tr}=10\,Rg$. In this regime, the importance of relativistic effects will decrease, and $R$ will converge to the non-relativistic limit \citep{2014MNRAS.444L.100D}. Therefore, the Newtonian treatment used in this work is sufficient for studying the qualitative dependence of $R$ on coronal outflow velocity and disk truncation radius.

\subsection{Intrinsic lag and zero-crossing frequency\label{sec:Intrinsic lag}}

The lag-frequency spectra of BHXRBs usually show hard lags at low Fourier frequencies and soft lags at high Fourier frequencies \citep{2017MNRAS.471.1475D, 2019Natur.565..198K, 2021A&A...654A..14D}. In our model, the low-frequency hard lag is mainly produced by inward propagation of accretion-rate fluctuations. Our model also includes intrinsic hard lags arising from the additional light travel time associated with successive Compton-scattering orders in the corona. However, for the coronal sizes adopted here, these scattering-induced lags are shorter than the propagation lags and therefore do not dominate the low-frequency variability. Since the propagation timescale is much longer than the light-crossing time between the disk and the corona, this process dominates the lag at low frequencies. High-frequency fluctuating signals usually originate from the inner radius of the disk, which implies that the time scale for the propagation of fluctuations is very short. Therefore, reverberation lags become more important at high frequencies.

In the following discussion, we use the zero-crossing frequency $\nu_0$, rather than the absolute lag amplitude, as the main timing diagnostic. The measured soft-lag amplitude can be diluted by the mixture of spectral components in each energy band, since both the reflection-dominated and continuum-dominated bands contain contributions from multiple emission components \citep{2014A&ARv..22...72U, 2015ApJ...814...50D}. Previous studies have shown that $\nu_0$ is less sensitive to this effect than the lag amplitude \citep{2015ApJ...814...50D, 2018MNRAS.478..971M, 2021A&A...654A..14D}. For a static corona, $\nu_0$ can therefore be used as a useful proxy for the characteristic reverberation timescale and, consequently, for the disk-corona distance.

However, this interpretation has a limitation in the presence of a fast outflow. For an extended corona, hard photons generally undergo more scatterings than soft photons, so the direct Compton components can have an intrinsic hard lag due to light travel delays between successive scattering orders, even at relatively high frequencies. In the simulations shown in Fig.\ref{fig:lag}, this intrinsic continuum lag is only about 1-2 ms for the selected energy bands and is not large enough to remove the soft lag when the reflection component is strong. Instead, it mainly dilutes the soft-lag amplitude. When $\beta$ becomes larger than about 0.5, the reflection component is strongly suppressed by relativistic beaming, and even the 0.1-0.3 keV band is no longer purely dominated by reflection. As a result, the observed decrease of $\nu_0$ with increasing $\beta$ mainly reflects the reduced contribution of reflection, rather than the change of intrinsic light-travel distance between the corona and disk. Therefore, for $\beta \gtrsim 0.5$, $\nu_0$ should not be directly interpreted as the intrinsic reverberation timescale.

This distinction is important for interpreting observations. In a static-corona or slowly outflowing-corona model, a change in $\nu_0$ mainly traces a change in geometry \citep{2021A&A...654A..14D, 2025ApJ...985..258Z}. In a fast outflowing corona, however, a shift in $\nu_0$ can be driven by the changing relative strength of the reflected and Comptonized components, even when the geometry is fixed. Therefore, timing diagnostics must be interpreted together with the reflection fraction and spectral slope.

\subsection{$R-\Gamma$ correlation and the evolution of corona\label{sec:correlation}}

The correlation between the reflection fraction $R$ and the photon index $\Gamma$ has been reported in both BHXRBs and AGNs \citep{1999A&A...352..182G, 2010LNP...794...17G, 2023RAA....23g5005D, 2023ApJ...945...65Y}. In most BHXRB observations, a larger $R$ is associated with a softer spectrum, i.e., a larger $\Gamma$. Two main physical scenarios have been proposed to explain this relation. In the truncated-disk scenario, the disk inner radius decreases as the source softens. The disk then subtends a larger solid angle as seen by the corona, so more coronal photons irradiate the disk and $R$ increases \citep{1997ApJ...489..865E, 2007A&ARv..15....1D, 2010LNP...794...17G, 2025ApJ...984..173F}. In the dynamic-corona scenario, a faster outflow beams photons away from the disk, reducing both the $R$ and the cooling by disk seed photons, which can also affect $\Gamma$ \citep{1999ApJ...510L.123B, 2001MNRAS.326..417M, 2021NatCo..12.1025Y}. Based on spectral information alone, these two scenarios can be difficult to distinguish.

\begin{figure}[h]
    \centering
    \includegraphics[width=\linewidth]{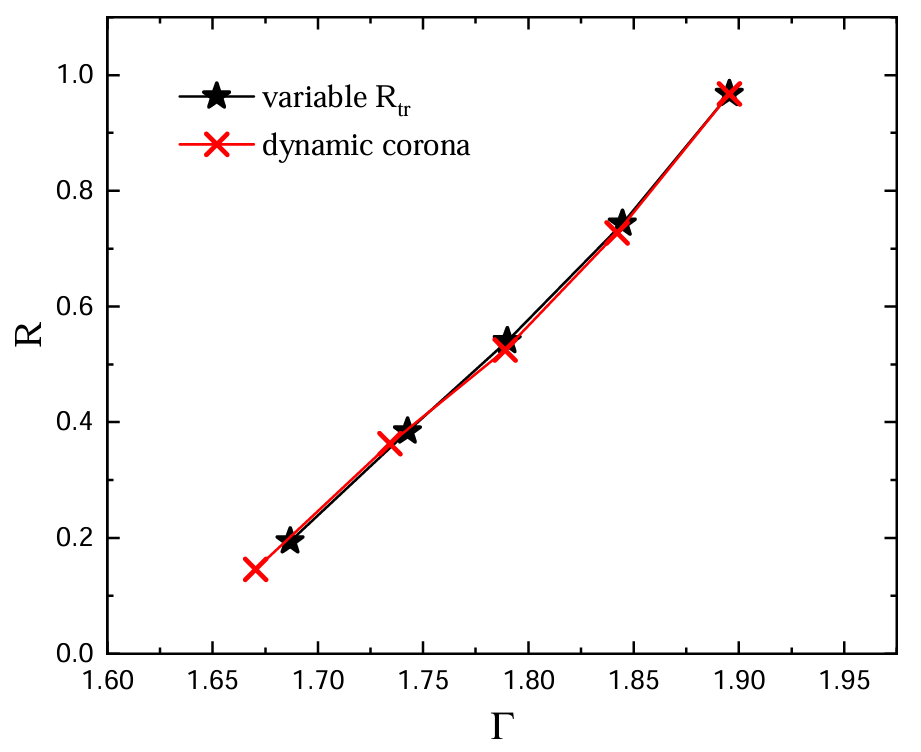}
    \caption{The $R-\Gamma$ correlation corresponding to the dynamic-corona scenario (red line) and the truncated-disk scenario (black line) given by our simulation. The geometric parameters of the corona are $a=120\,Rg$, $b=100\,Rg$. For the same coronal geometry, these two scenarios cannot be distinguished. For the dynamic-corona scenario, the red crosses from the upper right to the lower left respectively represent $\beta$ = 0.0 - 0.1 - 0.2 - 0.3 - 0.5. For the truncated-disk scenario, the solid stars from the upper right to the lower left respectively represent $R_{\rm tr}$ = 10 - 45 - 70 - 90 - 120 $Rg$.}
    \label{fig:Rgamma}
\end{figure}

\begin{figure}[h]
    \centering
    \includegraphics[width=\linewidth]{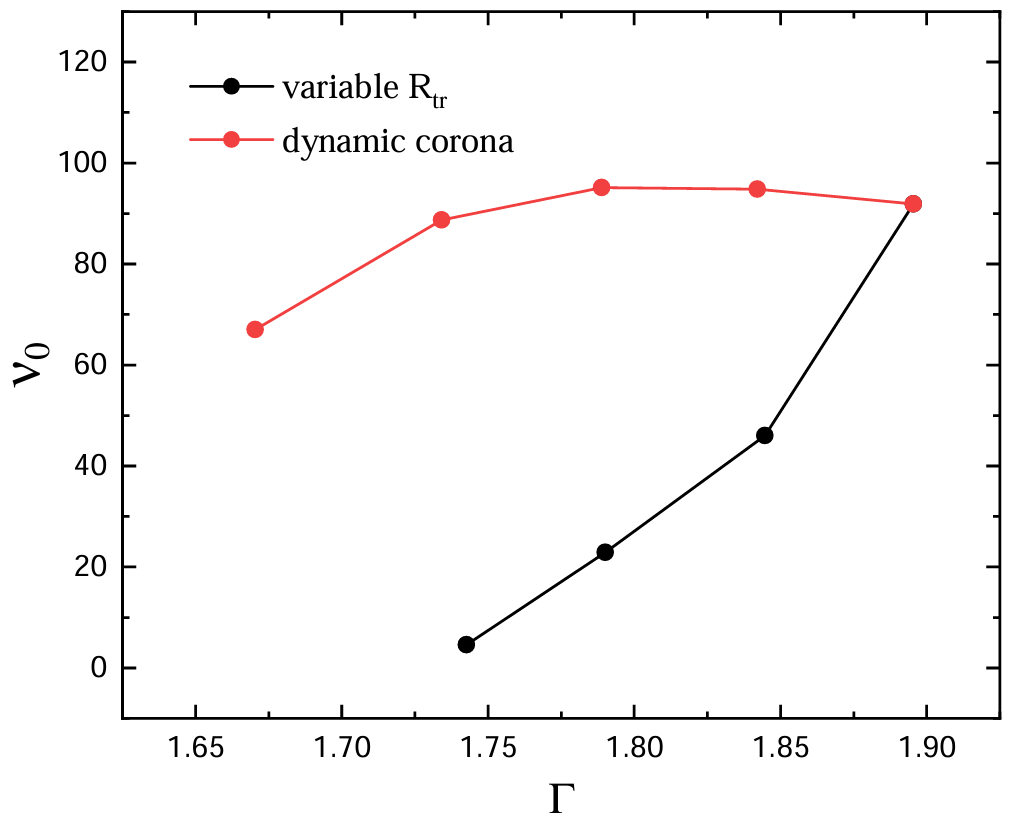}
    \caption{The $\nu_0-\Gamma$ correlation corresponding to the dynamic-corona scenario (red line) and the truncated-disk scenario (black line), where the settings of $\beta$ and $R_{tr}$ are the same as those in Fig.\ref{fig:Rgamma}. As $R_{tr}$ increases, $\nu_0$ significantly decreases, which is significantly different from the variation of $\beta$.}
    \label{fig:v0gamma}
\end{figure}

Fig.\ref{fig:Rgamma} shows the $R-\Gamma$ relation predicted by our model for the dynamic-corona scenario (red line) and the truncated-disk scenario (black line). For the same coronal geometry, both a changing outflow velocity and a changing disk truncation radius can reproduce a positive $R-\Gamma$ relation, as concluded in \cite{1999MNRAS.303L..11Z}. To break this degeneracy, we use the zero-crossing frequency $\nu_0$ as an additional timing diagnostic. Fig.\ref{fig:v0gamma} compares the $\nu_0-\Gamma$ relation produced by the two scenarios. When the coronal geometry is fixed and $\beta \lesssim 0.5$, increasing $\beta$ has only a weak effect on $\nu_0$. This is because the intrinsic disk-corona light travel distance does not change. By contrast, increasing $R_{\rm tr}$ in a static corona significantly decreases $\nu_0$, corresponding to a longer reverberation timescale. Similar behavior has been found in previous studies, where a contracted corona or a smaller $R_{\rm tr}$ shifts $\nu_0$ to higher frequencies \citep{2019Natur.565..198K, 2021A&A...654A..14D, 2023ApJ...953..191Y, 2025ApJ...985..258Z}. Therefore, if a source shows a low reflection fraction together with a relatively high $\nu_0$, the result favors a faster outflowing corona rather than a larger truncation radius.

\begin{figure*}[!p]
   \centering
   \includegraphics[width=0.85\linewidth]{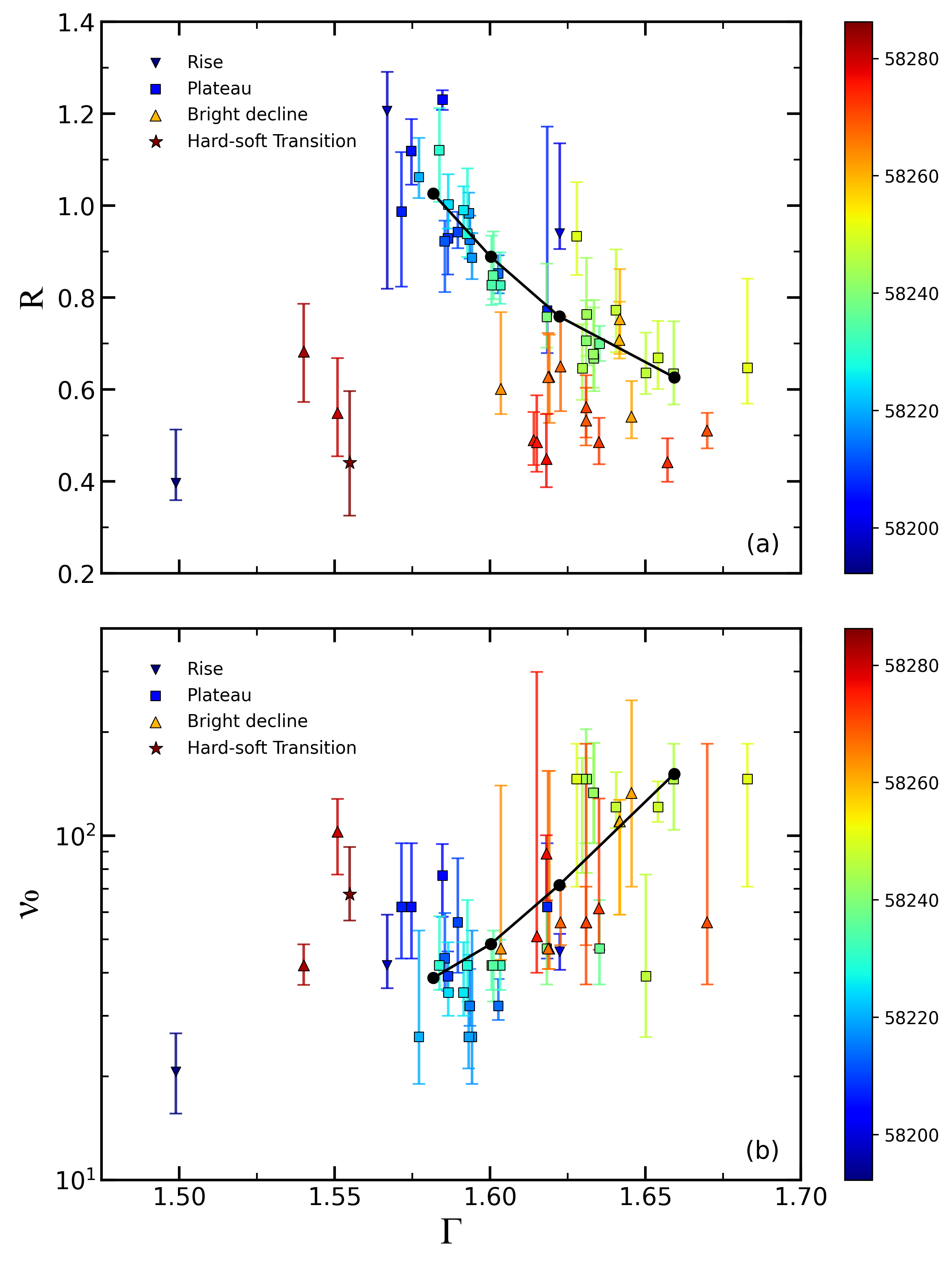}
   \caption{$R-\Gamma$ relationship figure (a) and $\nu_0$-$\Gamma$ relationship figure (b) of MAXI J1820+070 before the outburst enters the soft state. Lower triangles, squares, upper triangles and stars represent the four phases of outburst. The $R$ and $\Gamma$ are the best-fit result of constant*TBabs(diskbb + xillverCp + relxillCp) model. The data are taken from the Hard X-ray Modulation Telescope (Insight-HXMT) from March 12, 2018 (MJD 58190) to July 6, 2018 (MJD 58306) \citep{2014SPIE.9144E..21Z}. Due to the effective area of the LE detector in Insight-HXMT being one order of magnitude smaller than that of NICER \citep{2016SPIE.9905E..1HG,2021NatCo..12.1025Y}, which led to an insufficiently precise derivation of lag in the low-energy band. Therefore, we adopt $\nu_0$ as listed in Table E.1 of \cite{2021A&A...654A..14D}. The black curve represents a possible evolutionary track in which the corona contracts from $a=220\,Rg$, $b=200\,Rg$ to $a=70\,Rg$, $b=50\,Rg$, while $\beta$ increases from 0 to 0.15.}\label{fig:1820}
\end{figure*}

MAXI J1820+070 provides a useful application of this diagnostic. During the plateau phase, the source shows an unusual anti-correlation between $R$ and $\Gamma$, as shown in Fig.\ref{fig:1820}(a), which differs from the standard positive relation observed in many BHXRBs \citep{2023RAA....23g5005D, 2023ApJ...945...65Y}. A decrease in $R$ could, in principle, be caused either by an increase in $R_{\rm tr}$ or by an increase in $\beta$. However, the timing behavior helps distinguish the two possibilities. During the same phase, $\nu_0$ increases significantly and is positively correlated with $\Gamma$, which is shown in Fig.\ref{fig:1820}(b). If the decrease in $R$ were caused by disk recession, the increasing $R_{\rm tr}$ would instead predict a lower $\nu_0$, opposite to the observed trend. Moreover, some spectral and spectral-timing studies suggest that $R_{\rm tr}$ remains approximately constant during the plateau phase of MAXI J1820+070 \citep{2021A&A...654A..14D, 2025MNRAS.538.1143L}. Therefore, the decrease in $R$ during the plateau phase is unlikely to be driven mainly by an increasing truncation radius.

A more consistent interpretation is that the corona contracts while its outflow velocity increases. Coronal contraction in MAXI J1820+070 has been inferred from reverberation observations and has also been reproduced in previous models, including the double-lamppost model and the extended-corona model \citep{2019MNRAS.490.1350B, 2019Natur.565..198K, 2021A&A...654A..14D, 2025ApJ...985..258Z}. At the same time, an increasing outflow velocity beams more photons away from the disk and lowers the reflection fraction, as suggested by the jet-like corona interpretation of MAXI J1820+070 \citep{2021NatCo..12.1025Y}. This combination naturally explains the simultaneous increase of $\nu_0$ and decrease of $R$. As an illustrative example, we show an evolutionary track in which the corona gradually contracts while its outflow velocity increases in Fig.\ref{fig:1820}.

We note that the simulated and observed values of $\nu_0$ are not intended for a direct quantitative comparison, because they are derived using different energy bands. \cite{2021A&A...654A..14D} measured $\nu_0$ between 0.5-1.0 and 2.0-5.0 keV, whereas we adopt 0.1-0.3 and 2.0-5.0 keV in the simulations. This choice is necessary because, over a substantial part of our parameter space, the 0.5-1.0 keV band contains a significant, and sometimes dominant, contribution from the direct Comptonized emission, so that the corresponding lag does not provide a clean measure of the reverberation signal. The comparison in Fig.\ref{fig:1820} should therefore be interpreted in terms of the evolutionary trend rather than the absolute value of $\nu_0$. The specific coronal dimensions adopted in the illustrative track are thus not uniquely constrained, and the important result is that a contracting corona, together with an increasing $\beta$ that reduces $R$, can reproduce the observed directions of the spectral and timing evolution.

The situation is less clear during the rising phase and the bright decline phase. In the rising phase, the increase in $R$ could be explained either by a decreasing $\beta$ together with coronal contraction or by a decreasing $R_{\rm tr}$. \cite{2021A&A...654A..14D} suggested that the disk inner radius decreases during the rising phase, which makes the truncation scenario difficult to exclude. In the bright decline phase, $\nu_0$ fluctuates over a broad range and does not show a monotonic trend. Therefore, timing information alone does not provide a unique distinction between the two scenarios during this period. If $R_{\rm tr}$ remains approximately constant, as suggested by \cite{2021A&A...654A..14D} and \cite{2025MNRAS.538.1143L}, then the observed variation in $R$ may still be mainly related to changes in $\beta$. More detailed spectral-timing modeling is required to test this interpretation.

\subsection{Polarization\label{sec:polarization}}

X-ray polarization provides an independent probe of the geometry and dynamics of the corona and can therefore complement reverberation measurements. Reverberation mainly constrains the characteristic distance between the corona and the disk, whereas polarization is sensitive to the geometry, orientation, and bulk motion of the Comptonizing region \citep{2022Sci...378..650K, 2022MNRAS.516.5907M, 2024ApJ...968...76I}. Different coronal geometries can produce different polarization degrees and polarization angles, providing an additional way to distinguish between radially extended and vertically extended coronae. We note that the fiducial coronal geometry adopted in Table \ref{tab:paramenters} may produce a lower polarization degree than that observed in some hard-state X-ray binaries. The relatively large vertical extent of the corona is required in our model to intercept sufficient disk seed photons and reproduce the observed photon index. In future work including general relativistic light bending, a smaller corona may provide a similar Comptonization efficiency while producing different polarization signatures.

Coronal outflow can further modify the polarization signal through relativistic aberration and the anisotropic angular distribution of scattered photons. Previous calculations have shown that the polarization properties depend sensitively on the bulk velocity of the corona \citep{1998ApJ...496L.105B, 2023ApJ...949L..10P}. In particular, outflowing corona models have been used to interpret the X-ray polarization observed in Cygnus X-1, and moderately relativistic jet-like or conical coronae can reproduce its observed polarization characteristics \citep{2023ApJ...949L..10P, 2024MNRAS.528L.157D}. These results suggest that combining reverberation and polarization measurements may provide stronger constraints on both the spatial extent and dynamics of the corona. In future work, we will extend the present Monte Carlo model to calculate the polarization signatures of extended outflowing coronae.

\section{Conclusions} \label{sec:conclusions}

In this work, we develop a Monte Carlo radiative-transfer model for an accreting BH surrounded by a truncated disk and an ellipsoidal outflowing corona. We investigate how the coronal outflow affects the spectral and timing properties. We also discuss the $R-\Gamma$ relation and the possible evolution of MAXI J1820+070 during the plateau phase. Our main conclusions are summarized as follows:

(1) An outflowing corona is intrinsically anisotropic. Due to relativistic beaming, Comptonized photons are preferentially directed along the outflow direction. This effect reduces the reflection fraction ($R$) and weakens the reflection component in the energy spectrum.

(2) For a fixed disk-corona geometry, increasing $\beta$ reduces the amplitude of the high-frequency soft lag. This decrease is mainly caused by the weakening of the reflected component, rather than by a change in the intrinsic light-travel distance between the corona and the disk. When $\beta \gtrsim 0.5$, the zero-crossing frequency $\nu_0$ no longer provides a direct measure of the intrinsic reverberation lag.

(3) For a given coronal geometry, both an increase in $\beta$ and an increase in the disk truncation radius $R_{\rm tr}$ can reduce $R$, and therefore both scenarios can reproduce a positive $R-\Gamma$ relation. However, their timing predictions are different. When $\beta \lesssim 0.5$, $\nu_0$ is only weakly affected by the outflow velocity. In contrast, increasing $R_{\rm tr}$ significantly decreases $\nu_0$. Thus, the joint behavior of $R$, $\Gamma$, and $\nu_0$ can help distinguish between the coronal-outflow scenario and the disk-truncation scenario.

(4) The unusual $R-\Gamma$ anti-correlation observed in MAXI J1820+070 during the plateau phase can be naturally explained if the corona contracts while its outflow velocity increases. In our model, the decrease in $R$ is mainly attributed to the increase in $\beta$, while the increase in $\nu_0$ is caused by the coronal contraction. Within the adopted parameter space, the inferred $\beta$ of MAXI J1820+070 remains below 0.3.

These results show that coronal outflow can significantly affect both the reflection spectrum and the reverberation signal. Therefore, dynamical effects of the corona should be considered when using spectral-timing measurements to infer the geometry of the inner accretion flow. Future polarization simulations of extended outflowing coronae will provide an additional way to test this scenario.

\begin{acknowledgments}

This work is supported by the Natural Science Foundation of China (NSFC) 12322307, 12273026, 12673053 and 12361131579; by “the Fundamental Research Funds for the Central Universities”; Xiaomi Foundation / Xiaomi Young Talents Program; The data analysis in this paper have been done on the supercomputing system in the Supercomputing Center of Wuhan University. BDM acknowledges support from the Spanish MINECO grants PID2023-148661NB-I00, PID2022-136828NB-C44, and CNS2025-166335. We thank Alexandra Veledina for helpful discussions and constructive comments on the manuscript.

\end{acknowledgments}

\bibliographystyle{aasjournal}
\bibliography{references}

\end{document}